\documentclass{sig-alternate-10pt}

\usepackage{url}
\usepackage{subfigure}
\usepackage{graphics}

\newcommand{\comments}[1]{}

\begin{document}

\title{5G-ICN : Delivering ICN Services over 5G using \\Network Slicing}

\numberofauthors{1}
\author{
\alignauthor
Ravishankar Ravindran, Asit Chakraborti, Syed Obaid Amin, Aytac Azgin and Guoqiang Wang\\
\smallskip
\affaddr{Huawei Research Center, Santa Clara, CA, USA.}\\
\email{\large \{ravi.ravindran, asit.chakraborti, obaid.amin, aytac.azgin, gq.wang\}@huawei.com}\\
}


\maketitle

\begin{abstract}

The challenging requirements of 5G--from both the applications and the architecture perspectives--motivate the need to explore the feasibility of delivering services over new network architectures. As 5G proposes \comments{or As 5G proposals consider} application-centric network slicing, which enables the use of new data planes realizable over a programmable compute, storage, and transport infrastructure, we consider Information-centric Networking (ICN) as a candidate network architecture to \comments{help}realize 5G objectives. This can co-exist with end-to-end IP services that are offered today. To this effect, we first propose a 5G-ICN architecture and compare its benefits (\emph{i.e.}, innovative services offered by leveraging ICN features) to current 3GPP-based mobile architectures. We then introduce a general application-driven framework that emphasizes on the flexibility afforded by Network Function Virtualization (NFV) and Software Defined Networking (SDN) over which 5G-ICN can be realized. We specifically focus on the issue of how mobility-as-a-service (MaaS) can be realized as a 5G-ICN slice, and give an in-depth overview on resource provisioning and inter-dependencies and -coordinations among functional 5G-ICN slices to meet the MaaS objectives.  
\end{abstract}

\section{Introduction} \label{sec:introduction}
The key driving factors for 5G, which have been laid out in \cite{ngmn}, include: ($i$) support for high-density Internet-of-things (IoT) devices and services with very stringent end-to-end requirements (\emph{e.g.}, latency of $1-10$ms); ($ii$) support for very high throughput, with the average being $50$Mbps in all urban conditions, and peaking at $1-10$Gbps in ideal conditions; ($iii$) support for a new service class consisting of tactile applications that simultaneously carry low latency and high reliability requirements. Accordingly, a significant differentiator operators seek in 5G is the transition towards a service-centric infrastructure \footnote{By this we mean an infrastructure that is operated in a top-down manner using a service aware management, control and data plane.} that is also capable of fostering new business models between operators and the popular over-the-top (OTT) providers. These factors along with shift in communication patterns from connecting hosts to efficient dissemination of information~\cite{icnsurvey} considering security and mobility requirements motivate the need to evaluate new network architectures (other than the currently applied IP networking). 

Current research efforts on 5G network architecture adopt two different views:
\begin{itemize}
\item The first view, as considered in \cite{metis}, proposes a 5G architecture with focus on the evolved radio access network (RAN), while preserving 4G's core network architecture but over a flexible NFV/SDN-based infrastructure. However, adopting 4G network architecture to 5G also means to inherit the drawbacks of the current IP architecture, with respect to: ($i$) complex core networking based on tunnelling technology to support mobility, ($ii$) security challenges leading to high signalling costs, ($iii$) lack of multihoming support, ($iv$) a network infrastructure that does not leverage the agility of cheap computing and storage resources in the transport infrastructure.

\item The second view, as considered by ITU's 5G focus group FG-IMT2020 \cite{imt2020}, acknowledging the heterogeneous service requirements, discusses the benefits of architectures like Information-centric Networking (ICN) with inherent support for features like name-based networking, storage, computing, security, and mobility. This is made feasible in 5G by the proposed network softwarization and the ability to slice the raw transport, compute and storage resources, among multiple services~\cite{ngmn}. Within the context of such 5G architecture that is driven by a network slicing framework, ICN can be realized as a slice, over which the services can be delivered.
\end{itemize}


In this article, adopting the second view, we discuss a 5G-ICN architecture based on the NFV/SDN-framework to realize a top-down service centric platform, in which the ICN-based service delivery platform becomes a natural extension of the cloud into the infrastructure. This is made possible as: ($i$) ICN allows compute, storage and network virtualization on the same platform, and ($ii$) an ICN-based service delivery can orchestrate complex service-logic execution by service-function placement and content processing at the extreme edges of the network (\emph{e.g.}, Base stations) while being extendible to commodity utilities (\emph{e.g.}, lamp posts or traffic lights). 
Furthermore, as ICN has proven its usefulness in constrained and ad hoc infrastructures \cite{emmnuel,vanet}, it also represents an ideal platform to deliver unified IoT services \cite{icniot} over the 5G framework. We exemplify the benefits of such a service platform by considering the case of delivering mobility-as-a-service (MaaS) over a converged programmable infrastructure which enables network slicing.

The remaining sections are laid out as follows. {Section~\ref{sec:icn}} provides a brief introduction to ICN, followed by discussion on a 5G-ICN architecture in {Section~\ref{sec:5gicnarch}}, with focus on accommodating IoT services and applications with high bandwidth requirements. Here, we explain the features enabled by this architecture considering the current 3GPP systems, and various 5G-ICN deployment scenarios. {Section~\ref{sec:slicing}} introduces a generalized network slicing framework, over which both IP and 5G-ICN services can be delivered. 
In {Section~\ref{sec:MaaS}}, we present a use case study of realizing the 5G-ICN architecture over a network slicing framework, where we discuss how mobility-as-service (MaaS) can be realized. We emphasize on how to bootstrap different ICN network and service slices and their interactions to achieve the MaaS objectives. We present our final remarks in Section~\ref{sec:conclude}. 

\comments{
\begin{figure}[ht]
\centering
\includegraphics[scale = 0.6]{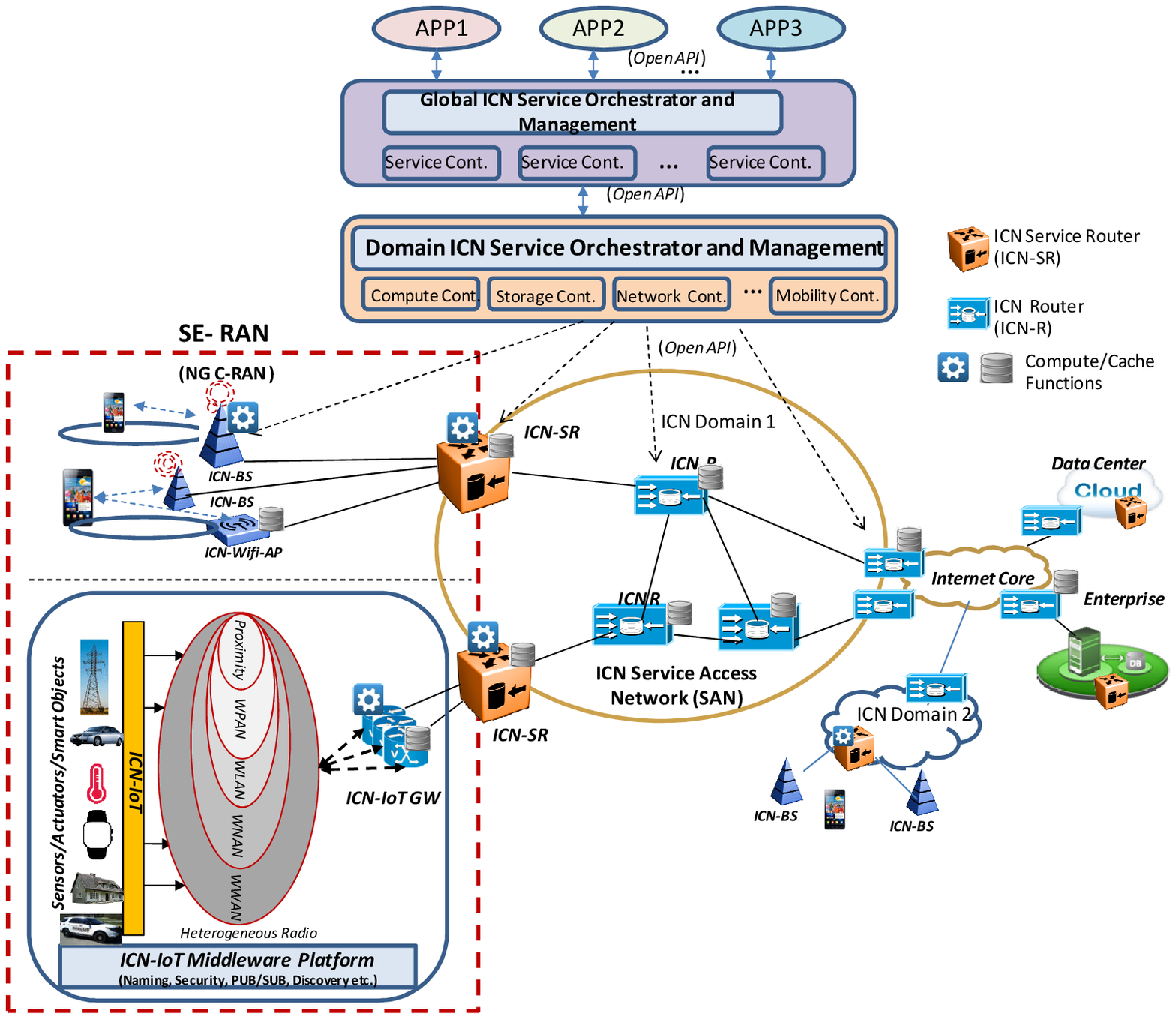}
\caption{5G-ICN architecture.}
\label{fig:5gicn}
\vspace{-.1in}
\end{figure}
}
\comments{
\begin{figure*}[htb]
    \begin{minipage}[b]{0.5\textwidth}
        \centering
        \subfigure[\small 5G-ICN architecture.]
        {
            \includegraphics[scale = 0.5]{figures/5gicn-eps}
            \label{fig:5gicn}
        }
    \end{minipage}
     \begin{minipage}[b]{0.5\textwidth}
        \centering
        \subfigure[\small ICN service slicing.]
        {
            \includegraphics[scale = 0.5]{figures/icn-slicing-eps}
            \label{fig:icnslicing}
        }
    \end{minipage}

    \caption{5g-ICN architecture and ICN service slicing.}
     \label{fig:networkslicingarch}
    \vspace{-.1in}
\end{figure*}
}

\section{Information-centric Networking}\label{sec:icn}
Information-centric Networking(ICN)~\cite{icnsurvey} is a result of various future network architecture research pursued in various parts of the world which enables features such as: (\emph{i}) name based networking of resources corresponding to contents, services, devices and network domains; (\emph{ii}) session-less transport through per-hop name resolution (of the requested resource), which also enables 5G-targeted features such as mobility, multicasting and multi-homing; (\emph{iii}) exploiting transport embedded compute-storage resources that are virtualizable among heterogenous services; (\emph{iv}) network layer security, which allows one to authenticate user requests and the returned content objects, thereby allowing location-independent caching and computing as desired by ICN applications and infrastructure providers; (\emph{v}) suitability to both adhoc- and infrastructure-based IoT environments, where the information-centric nature of IoT applications matches with what the ICN infrastructure offers.

We next discuss an ICN based 5G architecture capable of leveraging these features to enable a service-oriented network architecture.

\begin{figure}[ht]
\centering
\includegraphics[scale = 0.55]{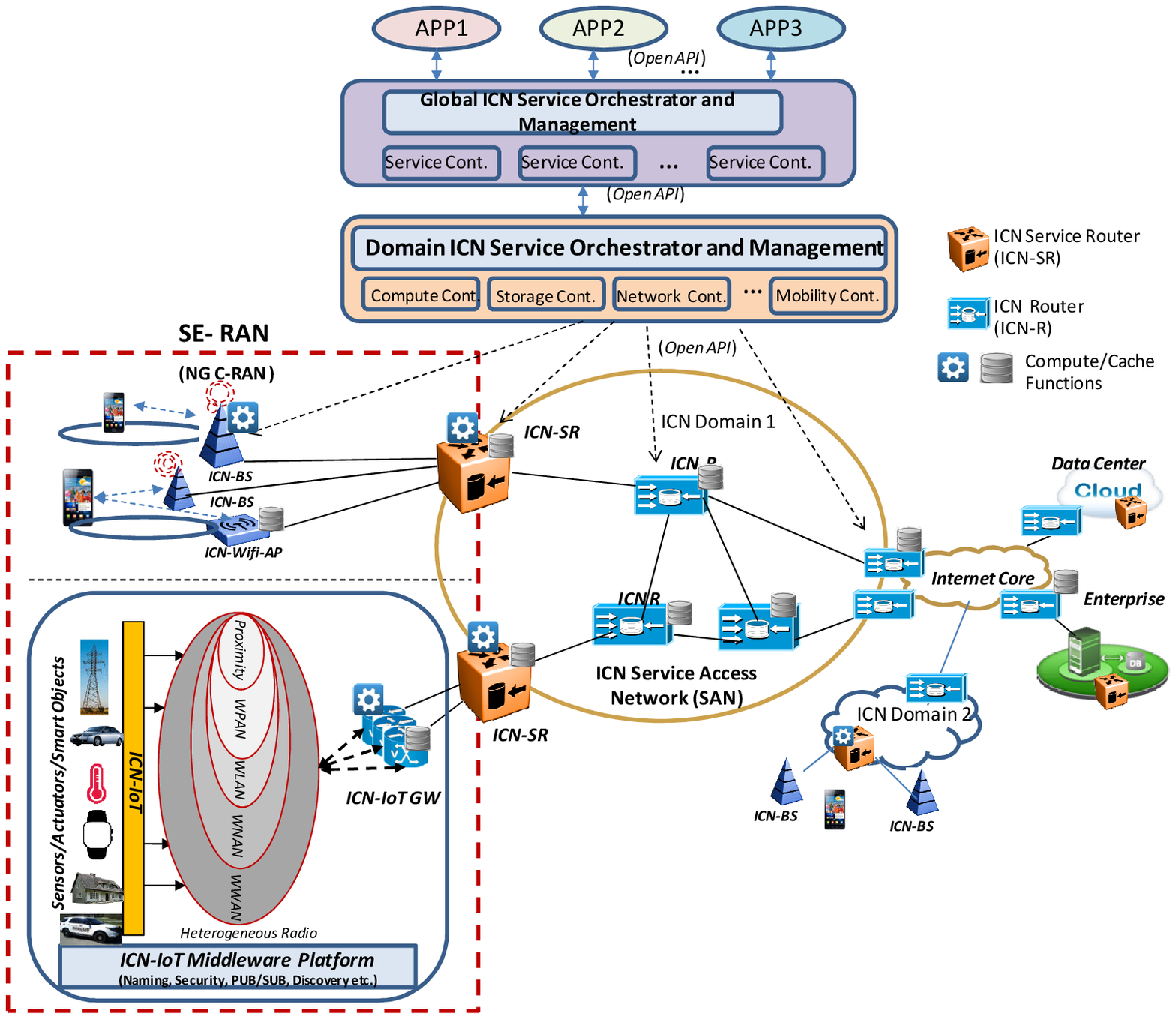}
\caption{5G-ICN Architecture.}
\label{fig:5gicn}
\vspace{-.1in}
\end{figure}

\section{5G-ICN Architecture}  \label{sec:5gicnarch}

5G presents a great opportunity for introducing new network architectures to address services requirements that are difficult to be satisfied with the current IP networking. The need for a new network architecture can be justified based on the following important benefits: (\emph{i}) to address the issue of having a single protocol that can handle mobility and security instead of having a diverse set of IP-based 3GPP protocols (as is the case for the current cellular systems), (\emph{ii}) to serve as a unifying platform with the same L3 APIs to integrate heterogenous radios (such as Wifi, LTE, 3G) and wired interfaces over which devices and services connect to the network, and (\emph{iii}) to converge computing, storage and networking over a single platform, which improves the flexibility of enabling virtualized service logic and caching functions anywhere in the network (especially for the access segment). All of these benefits can be achieved by a 5G architecture based on ICN (\emph{i.e.}, 5G-ICN).

In the data plane, 5G-ICN is capable of realizing a flat architecture without specialized gateways, as shown in Figure~\ref{fig:5gicn}, where applications and devices seek connectivity through the RAN to ICN gateways (resulting in service-enabled RAN or {SE-RAN} and co-located with cloud RAN implementation hence called as next generation cloud RAN or {NG-RAN}).
In  Figure~\ref{fig:5gicn}, we refer to the edge ICN routers as ICN service routers (ICN-SR), as these nodes are equipped with additional compute, storage and bandwidth resources to be shared among services. The inherent ICN capabilities (such as in-network caching and computing, and multi-homing features) enable 5G-ICN to naturally offer support for high bandwidth applications. Similarly, the suitability of ICN to support IoT applications \cite{middleware} (due to exploiting, for instance, naming, device-to-device communications, contextual networking, and self-X features such as configuration, management and healing) allows IoT services to be efficiently delivered using the same protocol infrastructure. The 5G-ICN segment for IoT services can be supported with distributed middleware service functions over ICN (such as device and service discovery, naming, context processing and pub/sub service) features required for the IoT systems. 

In control and service planes, considering the service-oriented networking requirement as stated by \cite{ngmn}, 5G-ICN naturally lends itself to service virtualization through a global service orchestrator. This can be realized through a logically centralized service plane, which abstracts resources from domain level orchestrators that monitor, manage and abstract ICN infrastructure resources within each domain. To support data plane virtualization, we consider sharing cache, compute and networking resources within ICN routers among multiple services using compute virtualization such as virtual network functions (VNF)\footnote{\emph{Virtual network functions} are virtual machines (VM) or containers required to support specific logical functions of the network such as IP/ICN forwarding, fronthaul/backhaul RAN processing and generic middlebox functions.}, P4 framework~\cite{P4}, or logical partitioning of a physical or software ICN router (similar to VPN technologies over IP today). 

Additionally, 5G-ICN can meet several requirements, which do not exist in current cellular network architectures such as LTE~\cite{3gpp-lte}, and these include:

{\textbf{Naming}}: 
Applications today conflate IP addresses as identifiers, hence it becomes difficult to support session mobility or achieve multihoming in an IP architecture. Unlike this, ICN applications bind to persistent names that are used to identify hosts, contents or services. Naming resources insulates applications from any kind of host mobility or even service mobility, as ICN layer handles the mapping from the high-layer application identifiers to locators. ICN allows different application-centric naming schemas, such as human-readable, self-certified or a hybrid one.  Self-certified names offer another desirable property, that is, authentication of hosts, services, devices or contents with minimal signalling cost. Such state can be managed at the BS to authenticate upstream or downstream transmissions. 

\textbf{Mobility}: As shown in Figure~\ref{fig:5gicn}, ICN enables a flat architecture, where mobility can be handled by the point-of-attachment (PoA) nodes, which in our architecture can be the ICN base station (ICN-BS) or the ICN-SR node integrating multiple radios. On the other hand, mobility in LTE~\cite{3gpp-lte} is handled by an orthogonal set of protocols. Specifically, in LTE, a per-user bearer tunnel state is created between the LTE-BS (eNodeB) and the Evolved Packet Core(EPC) containing the Service Gateway (S-GW) and the PDN gateway (P-GW), over which the UE's incoming/outgoing traffic tunnels, as the UE handovers from one eNodeB to another. As the amount of tunnel state required to handle mobility in the data plane is proportional to the number of UEs, signalling overhead increases with host dynamicity.


\textbf{Security}: In current LTE systems, it typically takes $60$ms for a device to go from idle state to active state before sending or receiving any data \cite{ltelat}, which is due mostly in part to {UE} authentication and signalling of the bearer paths--between eNodeB, S-GW and P-GW--for the UE traffic. In the case of ICN, application APIs for Interest/Data traffic\footnote{Note that, the Interest/Data primitives belong to CCN/NDN~\cite{ccn}, but are similar to that of other ICN protocols such as MobilityFirst~\cite{mobilityfirst}.} bind identity information to enable security features such as content integrity and provenance validation. Also the primitives associated with ICN traffic can be contextualized with additional security attributes such as device or user identity, which can be subjected to in-network security verification.

\textbf{Reliability}: The features inherently offered by ICN, such as sessionless store-and-forward operation, per-hop name resolution\footnote{This applies to both Interest and Data.}, per-hop congestion control, multi-homing, replicated caching, and multi-path routing, allow quick and painless recovery from congestion scenarios or link failures in manners not achievable with IP networking.

\textbf{Efficiency}: ICN offers efficiency at every level, from data plane to control and management planes. Data plane efficiency is achieved through replicated computing, caching and storage in the network, thereby reducing costs associated with upstream bandwidth, storage and computation uses. Control plane efficiency is achieved through name-based networking that allows different modes of control plane techniques based on the networking environment, such as flooding mechanism in the case of device-to-device communications for ad hoc scenarios, or leveraging traditional routing mechanisms to enable unicast, multicast or anycast in the infrastructure. Management plane efficiency is achieved by several Self-X features that ICN provides, such as per-hop congestion control, multi-path routing, and minimal overhead for bootstrapping.

\textbf{Contextual Communication}: ICN APIs are service-centric and contextualized by nature. For instance, consumers can make requests for a content using optional contextual metadata (\emph{i.e.}, location, device, or criticality of information request or response), which can then be subjected to in-network processing at the ICN routers or through overlaid virtual service functions (VSF)\footnote{\emph{Virtual service functions} are responsible for executing ($i$) the specific service logics related to the services, or ($ii$) the generic ones to aid with service discovery and naming services.} for further processing. In doing so, these contextualized requests can be satisfied at the network edge, which can improve UEs' QoE to meet the service objectives.

\begin{figure}[t]
\centering
\includegraphics[scale = 0.5]{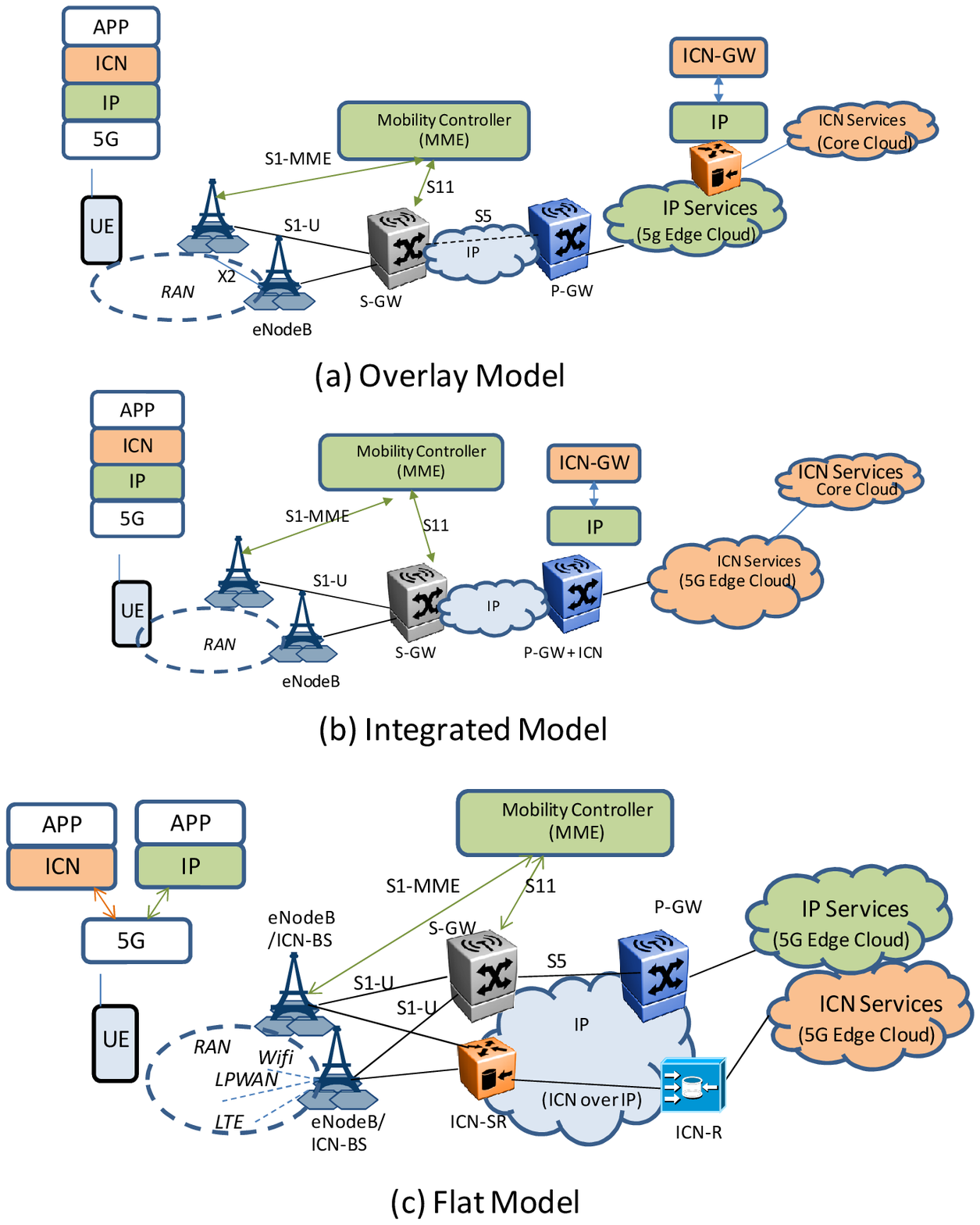}
\caption{ICN deployment models.}
\label{fig:icndeploy}
\vspace{-.1in}
\end{figure}

\subsection{5G-ICN Deployment Models}

We consider three possible 5G-ICN deployment scenarios, which are shown in Figure~\ref{fig:icndeploy} and discussed next using the LTE \cite{LTEMob} architecture as a reference.

\textbf{Overlay Model} is shown in Figure~\ref{fig:icndeploy}(a), in which ICN becomes an overlaid service over the current IP infrastructure. Despite being overlaid, ICN can still be realized as a service platform that is managed by the operator, while offering caching and compute benefits through the edge and core cloud infrastructure (which can serve a large regional geography). 

\textbf{Integrated Model} is shown in Figure~\ref{fig:icndeploy}(b), in which ICN becomes tightly integrated with the core mobile network infrastructure. This model assumes an explicit control and management plane to relay ICN PDUs from UE over 5G to the P-GW, which hosts the ICN router. Hence for an ICN service slice, ICN service flows can choose to route to different ICN P-GWs based on the service requirements. In this scenario, mobility is handled by the underlay 5G protocol, but the benefits of caching and computing are distributed within the core infrastructure and closer to UE. In this model the ICN channel could be enabled using the same control and signalling infrastructure of the LTE.

\textbf{Flat Model} is shown in Figure~\ref{fig:icndeploy}(c), in which the 5G-ICN architecture (discussed in Section~\ref{sec:5gicnarch}) is integrated within the network to take advantage of all the benefits offered by 5G-ICN. As ICN integrates security and anchorless mobility, other overlay protocols are not required in the ICN infrastructure. Even if, control functions such as  MME, Home Subscriber Service (HSS) or Policy and Charging Rule Functions (PCRF) are required, they can be adapted to the ICN network after accounting for the features enabled by ICN in the network layer. 

With the above view of the 5G-ICN architecture, we discuss how 5G-ICN can be realized in a framework capable of offering network slicing service.

\begin{figure}[t]
\centering
\includegraphics[scale = 0.5]{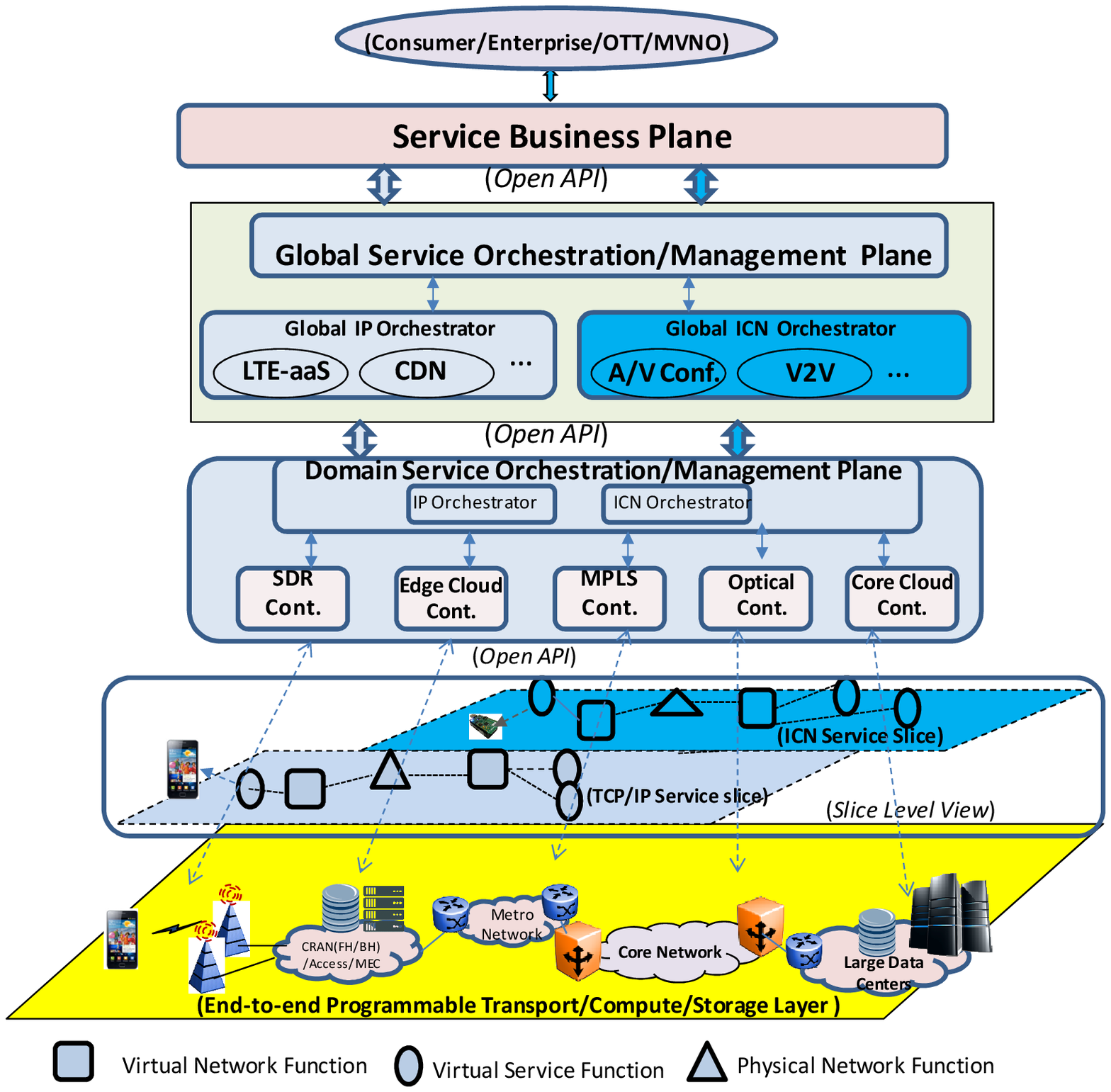}
\caption{Network slicing framework.}
\label{fig:slicingarch}
\end{figure}

\section{Network slicing architecture} \label{sec:slicing}
Slicing a 5G network on an end-to-end basis, which spans multiple technology domains and include the user equipment (\emph{UE}) resources, aims to support a diverse set of applications with different service requirements (\emph{e.g.}, latency, bandwidth and reliability) using a common resource pool consisting of compute, storage, and bandwidth resources. Figure~\ref{fig:slicingarch} shows a generic network slicing architecture with capability to create IP and 5G-ICN service slices. The framework has the following five functional planes (FPs):

\textbf{\texttt{FP1} - {Service Business Plane}} forms the interface between the external 5G service users and the softwarized infrastructure that helps realize connectivity and user-centric services in a dynamic manner. \texttt{FP1} exposes various service APIs, which the network is capable of delivering through a formal intent model, along with service management and monitoring functions. The business plane APIs can be realized as high level abstraction language, through which the service expresses, ($i$) what it wants to accomplish and ($ii$) the network services required to accomplish it. For instance, service input can include service type, demand patterns, and requirements on SLA/QoS/QoE and network services such as reachability, security, mobility, multicasting and storage. \texttt{FP1} converts these requirements into information-and-data models as required by \texttt{FP2}.

\textbf{\texttt{FP2} - {Service Orchestration and Management Plane}}, upon receiving the service requests from \texttt{FP1}--with explicit information on the narrow waist to use for service delivery--communicates the service requirements to the respective IP/ICN global service orchestrators for their execution.  

\textbf{\texttt{FP3} - {IP/ICN Global Orchestrator}} realizes IP and ICN services, by leveraging the already existing slices (if necessary). In Figure~\ref{fig:slicingarch}, IP and ICN service orchestrators are logically separated, as the network and services operate on different data, control and service plane APIs. \texttt{FP3} interfaces with domain controllers to virtualize compute, storage and network resources to meet the service requirements, with the help of the following functions: ({\emph{i}) translating service requirements to resource requirements in the data plane, and identifying different VNF/VSFs required to support the given service, while generating a slice context for service, control and data plane management; ({\emph{ii}) monitoring compute, storage and network resources at the edge and core clouds, and transport segments \footnote{In the case of ICN, compute and storage resources become part of the infrastructure}; ({\emph{iii}) keeping an abstract view of the physical (topology) resources and its mapping in the context of multiple slices; ({\emph{iv}) interfacing with technology specific domain controllers (in \texttt{FP4}), to enforce the rules determined by the service orchestrator; ({\emph{v}) handling global life cycle management of the VNF/VSFs, failure management and network reliability based on the service layer agreement (SLA) requirements.

\textbf{\texttt{FP4} - {Domain Service Orchestration and Management}} support orchestration of IP and ICN services within domains. As end-to-end network segments will be comprised of multiple domains with differing technologies, ranging from 4G/5G RAN to Optical/MPLS transport domains, and from edge to central cloud resources, each of these domains will be governed by its own local network, compute, and storage controllers. ICN controllers is realized as sub-controllers in the ICN relevant domains. Functions such as domain slice SLA management, VNF/VSFs life cycle, failure management are also handled by the domain controllers in coordination with the \texttt{FP3} function.

\textbf{\texttt{FP5} - {Infrastructure Plane}} distributed among multiple domains and managed by \texttt{FP4},  enables the service rules in an end-to-end manner, spanning the UEs, RAN and access networks, heterogenous transport segments, edge and central clouds.

In the context of multiplexed IP/ICN flows, transport plane should be able to differentiate among these flows to provide the appropriate resource guarantees. In the case of ICN, finer flow-level service differentiation depends on the capability of network understanding the ICN primitives (\emph{i.e.}, name based flows, resources to manage the multicast state \footnote{This includes both multiple user requests for the same content, or pushing a content to multiple receivers.}, request forwarding and software-defined cache management policies, in-network computing and context processing rules, and, QoS and queue management policies).

Software-defined radio (SDR) allows the realization of flexible and elastic MAC/PHY layers to cater to a diverse set of services, such as low-power IoT and high-bandwidth video applications. 
Various middlebox, data plane, control and service functions can be realized over the generic Intel-x86  infrastructure in the form of VNF/VSFs. Data plane can be based on virtual network overlays or deeply programmable hardware based on the P4 or OpenFlow technologies to multiplex IP/ICN service flows to achieve QoS isolation and line-rate switching. UE programmability enables mapping IP/ICN flows, \emph{e.g.} using host virtual switch, to appropriate radio slices and then handing over the service flows at the Base Station(BS) to the appropriate service slice.

Having provided a broad discussion of the network slicing framework capable of delivering both IP and ICN services, we next discuss the realization of mobility-as-a-service as a 5G-ICN slice and its functional interaction with other 5G-ICN service slices.


\begin{figure}[t]
\centering
\includegraphics[scale = 0.5]{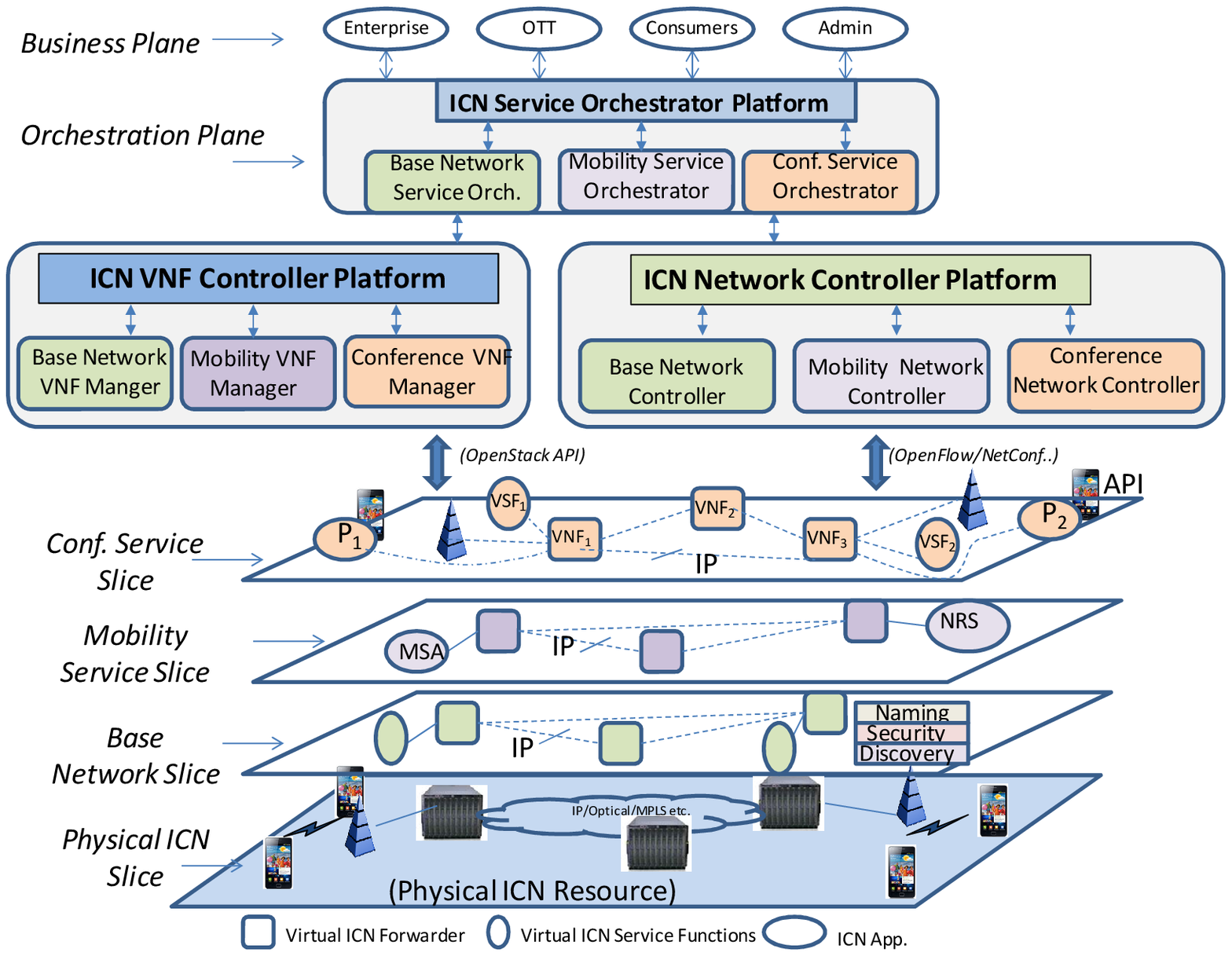}
\caption{Mobility-as-a-service realization.}
\label{fig:MaaS}
\vspace{-.1in}
\end{figure}

\section{5G-ICN Mobility-as-a-Service} \label{sec:MaaS}
In this section, as a case study of realizing 5G-ICN within a network slicing framework, we discuss mobility-as-a-service (MaaS) and how other service slices can use the APIs exposed by MaaS to enable mobility service to its dynamic entities. We consider the following objectives for MaaS:
\begin{itemize}
\item \textsf{On-demand mobility} allows ICN names to be dynamically (de-)registered for the mobility service, and when registered, all the flows for that name are provided with mobility support.
\item \textsf{Minimal session disruption} is needed to provide seamless mobility support to service flows within a mobility network slice, as a member UE moves from one PoA to another within the same slice. 
\end{itemize}
Even though the specific details of handling mobility in ICN vary depending on the protocol, the general principal remains the same, \emph{i.e.}, separating application name binding from the network address, which is also referred as the ID/Locator split and late-binding feature~\cite{aytacmob} that allows ICN PoA to redirect flows to UE's new PoA in a dynamic manner.

For this case study, we assume the ICN network is realized as a set of virtual entities, such as using a \emph{Container} technology, hence the state within a virtual ICN forwarders comprises only of the service states such as cached items and name reachability state that only remain for the lifetime of the virtual slice instance. 

\subsection{MaaS Operation}
We show the overall architecture that enables MaaS, which any service slice can leverage, in Figure~\ref{fig:MaaS}. This architecture is based on the discussion provided in Section~\ref{sec:slicing}. 
The application we assume is a video conferencing service, in which the participants join the conference randomly and can solicit audio/video/text content from other desired parties dynamically. Here, the global and domain service controllers, VNF/VSFs of each service expose the appropriate APIs that can be used by each other to achieve a service objective. We next explain how MaaS is realized considering the various stages of slice provisioning and interactions among the different slices:

\textsf{Step 1 - Base network slice bootstrap}: To support ICN service virtualization, first, we need to bootstrap the service functions that enable UE applications to discover and name the services, provide security functions, and connect them to the appropriate service gateway. We call this as the Base Network Slice, which is managed by the Base Network Slice VNF Orchestrator. ICN connectivity of this service slice among the various VNFs\footnote{The VNFs here are the virtual ICN forwarders.} and the UE is managed by the Base Network Slice Controller.

\textsf{Step 2 - Mobility network slice bootstrap}: As the mobility service is bootstrapped by the Mobility Service Orchestrator, two important service functions are enabled in the corresponding slice: (\emph{i}) Mobility Service Agent (MSA), which exposes APIs for the name resolution, and (\emph{ii}) Name Resolution Service (NRS), which maps the registered names to the corresponding locators in the network. For entities outside a domain, we assume the NRS to have APIs for inter-domain resolution. As NRS is a very critical component, the mobility compute and network controllers should ensure high availability for this service (see Mobility Network Slice in Figure~\ref{fig:MaaS}). 

\textsf{Step 3 - Creating a video conferencing slice}: As an external trigger for a video conference instance arrives from the business plane to the Conference Service Orchestrator, it maps the conference requirements (\emph{i.e.}, location information, number of participants at a physical location, device types, \emph{etc.}) to the ICN VNFs and VSFs with appropriate compute, cache and bandwidth resources to manage the expected traffic load. The Conference Service VNF Manager provisions the requested set of virtual ICN forwarders and service functions to support the conference session. The Conference Network Controller manages the connectivity between the virtual forwarders and service functions, and maps the dynamically arriving participants and their requests to appropriate VNFs for load balancing. Also, appropriate forwarding rules are pushed into the VNF instances to handle the service flows (see Conference Service Slice in Figure~\ref{fig:MaaS}).

\textsf{Step 4 - UE application bootstrap}: The ICN application at the UE discovers the service to connect to, through well known APIs available for service discovery (which is provided by the Base Network Service Slice). The discovery results in application receiving names, keys and trust information, and connecting to the appropriate slice gateway in the conference slice. For instance, in Figure~\ref{fig:MaaS}, UE's application instance, $P_{1}$, connects to the gateway $VNF_{1}$ when the UE joins the conference session.

\textsf{Step 5 - Enabling dynamic mobility}: Assume that an external trigger from the business plane requests for mobility service for the participants in a given conference slice instance. This request is first received by the Conference Service Orchestrator, which invokes the service APIs provided by the Mobility Service Orchestrator that pushes the request to the Mobility Network Controller, which sets the appropriate policy state (depending on the specific ICN protocol) in the MSA and the NRS within the mobility network slice. At the same time, Conference Service Orchestrator triggers the Conference Network Controller to register the mobile named entities of that slice to the NRS. The Conference Network Controller configures the ICN virtual forwarders to invoke resolution function to handle ICN flows bound to these mobile names. In short, for an incoming ICN request, the resolution request from the conference slice is passed to the MSA function in the mobility network slice, which then invokes the NRS for resolution to mobile participant's current location.

\textsf{Step 6 - Handling seamless mobility}: Late-binding mechanism (\emph{e.g.}, \cite{aytacmob} for NDN) can be used by the conference slice in the data plane to handle seamless mobility of the participants and to achieve minimal session disruption for the voice/video sessions handled by the conference slice. 

Note that, similar to the above scenario, mobility service can be disabled or enabled by a trigger in the business plane over any service slice.


\section{Conclusion}\label{sec:conclude}
In this article, we explored the feasibility of realizing future networking architectures, like ICN, under the network slicing framework proposed for 5G. We argued that, while ICN simplifies the network architecture, it can help meet the heterogenous service objectives leveraging the several desirable features of ICN. We examined a potential 5G-ICN architecture, explaining the features it enables and the possible deployment models. Finally, we studied how Mobility-as-a-Service can be dynamically enabled as a service slice considering a 5G-ICN framework and how other service slices could leverage it in a dynamic manner.

\small
\bibliographystyle{unsrt85}
\bibliography{paper-demo}

\begin{thebibliography}{10}

\bibitem{ngmn}
{NGMN 5G White Paper}.
\newblock [link]:
  \url{https://www.ngmn.org/uploads/media/NGMN_5G_White_Paper_V1_0.pdf}, 2015.

\bibitem{icnsurvey}
G.~Xylomenos, et~al.
\newblock A survey of information-centric networking research.
\newblock {\em IEEE Communications Surveys Tutorials}, 16(2):1024--1049, Second
  2014.

\bibitem{metis}
H.~Droste, et~al.
\newblock The metis 5g architecture: A summary of metis work on 5g
  architectures.
\newblock In {\em 2015 IEEE 81st Vehicular Technology Conference (VTC Spring)},
  pages 1--5, May 2015.

\bibitem{imt2020}
{ITU FG-IMT 2020, Network Standardization Requirement for 5G }.
\newblock [link]:
  \url{http://www.itu.int/en/ITU-T/focusgroups/imt-2020/
Documents/T13-SG13-151130-TD-PLEN-0208!!MSW-E.docx
  }, 2015.

\bibitem{emmnuel}
Emmanuel Baccelli, et~al.
\newblock Information centric networking in the iot: Experiments with ndn in
  the wild.
\newblock In {\em Proceedings of the 1st International Conference on
  Information-centric Networking}, ICN '14, pages 77--86, New York, NY, USA,
  2014. ACM.

\bibitem{vanet}
G.~Grassi, et~al.
\newblock Vanet via named data networking.
\newblock In {\em Computer Communications Workshops (INFOCOM WKSHPS), 2014 IEEE
  Conference on}, pages 410--415, April 2014.

\bibitem{icniot}
Yanyong Zhang, et~al.
\newblock Requirements and challenges for iot over icn.
\newblock In {\em draft-zhang-icnrg-icniot-requirements-01.txt,
  IETF/IRTF/ICNRG}, 2016.

\bibitem{middleware}
Yanyong Zhang, et~al.
\newblock Icn based architecture for iot.
\newblock In {\em draft-zhang-icnrg-icniot-architecture-00.txt,
  IETF/IRTF/ICNRG}, 2016.

\bibitem{P4}
Pat Bosshart, et~al.
\newblock P4: Programming protocol-independent packet processors.
\newblock {\em SIGCOMM Comput. Commun. Rev.}, 44(3):87--95, July 2014.

\bibitem{3gpp-lte}
3GPP-LTE.
\newblock 3gpp lte specifications.
\newblock In {\em http://www.3gpp.org/ftp/Specs/html-info/36-series.htm}, 2008.

\bibitem{ltelat}
D.~Singhal, et~al.
\newblock Lte-advanced: Handover interruption time analysis for imt-a
  evaluation.
\newblock In {\em Signal Processing, Communication, Computing and Networking
  Technologies (ICSCCN), 2011 International Conference on}, pages 81--85, July
  2011.

\bibitem{ccn}
Van Jacobson, et~al.
\newblock Networking named content.
\newblock In {\em Proceedings of the 5th international conference on Emerging
  networking experiments and technologies}, CoNEXT '09, pages 1--12, New York,
  NY, USA, 2009. ACM.

\bibitem{mobilityfirst}
Arun Venkatramani, et~al.
\newblock Mobilityfirst: A mobility-centric and trustworthy internet
  architecture.
\newblock In {\em ACM Sigcomm CCR, Volume 44, Number 3}, 2014.

\bibitem{LTEMob}
K.~Dimou, et~al.
\newblock Handover within 3gpp lte: Design principles and performance.
\newblock In {\em Vehicular Technology Conference Fall (VTC 2009-Fall), 2009
  IEEE 70th}, pages 1--5, Sept 2009.

\bibitem{aytacmob}
Aytac Azgin, Ravishankar Ravindran, and Guo-Qiang Wang.
\newblock A scalable mobility-centric architecture for named data networking.
\newblock In {\em ICCCN (Scene Workshop)}, 2014.

\end{thebibliography}

\end{document}